# MgB$_2$/Cu Racetrack Coil: Winding and Transport Testing

M.D. Sumption[1], M. Bhatia[1], M. Rindfleisch[2], J. Phillips[2],

M. Tomsic[2], and E.W. Collings[1]

[1]LASM, Materials Science and Engineering Department,

OSU, Columbus, OH 43210, USA

[2]Hyper Tech Research, Inc. Columbus, OH 43212, USA

**Abstract**

A racetrack coil using MgB$_2$/Cu strand has been fabricated and tested for transport current density at 4.2 K in self field. The monofilamentary strand was 1.0 mm OD and insulated with S-glass braid. Eighty turns of strand (42 m) were wound onto a stainless steel former with outer dimensions 25 cm x 10 cm x 0.86 cm. The resulting racetrack coil was heat treated at 675°C for 30 minutes in flowing Ar. The strand, with a superconducting fraction of 26%, occupied 49% of the total coil pack cross sectional area. The coil $I_c$ at 4.2 K and self field was 120 A (using a 1 µV/cm criterion). This led to a $J_{c,sc}$ (across the whole coil) of 6.12 x 10$^4$ A/cm$^2$, a $J_e$ in the wire of 1.59 x 10$^4$ A/cm$^2$, and an overall winding $J_e$ of 7.9 x 10$^3$ A/cm$^2$ at 4.2 K in self field. The *n*-values ranged from 17 to 56.

**Introduction**

Many groups now fabricate $MgB_2$ wires [1-13], powder-in-tube-processed (PIT) strands being favored for long length applications. Typical present day strands employ Fe or Cu, with perhaps Cu-Ni or monel as the outer sheath material. There are two main variants of PIT $MgB_2$ fabrication: ex-situ [1-4], and in-situ [6,10-13]. Each of these choices has advantages and disadvantages. Numerous efforts to develop $MgB_2$ strand are ongoing, and significant progress is being made in improving its basic properties such as transport $J_c$, upper critical field, and irreversibility field. Development of high quality multifilamentary (MF) strand geometries is now becoming important as well. Even now, however, the potential of $MgB_2$ in coil form is significant; it can fill an important niche as an inexpensive, lightweight conductor able to operate at temperatures of up to about 20 K.

Additionally, low AC loss versions of this conductor are being developed for use in high *dB/dt* environments. These conductors open up possibilities for transformers, motors, and generators, for the specific cases where 20 K operation is feasible. In particular, recent developments are leading towards light weight, superconducting exciter, stator, and rotor coils for certain aircraft motors that could operate in liquid hydrogen. The lighter weight coils, especially rotors, will enable the use of lighter weight support structures, and hence overall reductions in the specific weight of the power device. In the present work we are detailing our development of superconducting $MgB_2$ rotor coils (racetrack) for such applications.

**Background**

Soltanian, Dou, et al., [14] have made small solenoids coils from monofilamentary, 1 mm OD, in-situ based $MgB_2$/Cu wires. The strand, of total length 3 m, reached a transport critical current, $I_c$, of 72 A (1.3 x $10^5$ A/cm$^2$) at self field and 4 K. Bhatia et al. have measured 1 m lengths of $MgB_2$/Cu and $MgB_2$/Fe/monel strands wound into helical samples (ITER barrels) [15], reaching $10^4$ and $10^5$ A/cm$^2$ at 4.2 K and 4 T for Cu and Fe-based strands, respectively. A somewhat larger solenoid coil was made by Tanaka, Togano, et al. (Hitachi) [16] using a monocore, ex-situ powder-based, $MgB_2$/Ni tape. The coil required 10 m of wire (80 turns), was wax impregnated, and reached an $I_c$ of 105 A at 4.2 K. Machi and Murakami [17] fabricated a solenoidal coil using 3.5 m of monofilamentary, 0.5 mm OD, in-situ based $MgB_2$/Cu which gave 76 A at 4.2 K in self field (4.4 x $10^5$ A/cm$^2$). Fang, Salama, et al [18] fabricated a squat solenoid with a total wire length of 4 m wound from 1 mm square, monofilamentary, in-situ powder-based $MgB_2$/Fe. The coil had an $I_c$ of 185 A at 4.2 K and 1 T. A solenoid fabricated by Hyper Tech Research (reported by Hascicek et al [19]) required 20 m of wire. The round strand, 1 mm OD, of in-situ powder-based $MgB_2$/Cu achieved 278 A at 4.2 K and self field. Its 170 turns were sol-gel insulated. Serquis, Civale, et al. [20] fabricated a solenoid using 25 m of 1 mm OD wire of ex-situ $MgB_2$ powder in a stainless steel sheath; it achieved an $I_c$ of 350 A in self field at 4.2 K.

In this work, we report the fabrication and testing of a racetrack coil whose 42 m of wire achieved an $I_c$ of 120 A at 4.2 K and self field.

**Strand Fabrication**

The continuous tube forming/filling (CTFF) process was used to produce $MgB_2$/Cu composite strands. This process, as developed at Hyper Tech Research (HTR), begins with the dispensing of powder onto a metal strip as it is being continuously formed into a tube. The starting 99.9% Mg powders were 325 mesh, and the 99.9% B powders were amorphous, at a typical size of 1–2 μm. The powders were V-mixed and then run in a planetary mill before tube filling. In this case Cu strip was used to encapsulate the powders. After exiting the mill at a diameter of 5.9 mm the filled overlap-closed tube was inserted into a full hard 101 Cu tube and drawn to final size. For further details of this process see [10,11]. The $MgB_2$/Cu strand used in this work (designated HTR 475) was 1.01 mm OD and contained a superconducting fraction of 26%.

**Former Design, Strand Insulation, Coil Winding, Subsequent Heat Treatment**

A schematic of the of the racetrack-type coil former used for this work is shown in Figure 1. The former was fabricated from 316 SS, and had overall dimensions of 26 cm x 10.2 cm x 8.64 mm. The eighty turns that were wound onto this former required a total of 42 m of strand in 16 layers with 5 turns per layer leading to a racetrack coil pack 6.35 mm thick with inner dimensions of 21 cm x 5.7 cm and outer dimensions of 25.4 cm x 10 cm. The strand was insulated with single S-glass braid insulation (no binder) woven onto the strand. Its presence added 0.13 mm to the strand diameter. Two Cu pads were placed on the outside of the flange for current connections. A slot in the flange near the coil center allowed the wire to be brought out to the first Cu pad, a similar slot near the coil edge enabled the wire connection to the other pad.

During winding (with 4 lbs of back tension), the two 6 mm thick stainless steel side plates kept the flange from flaring out in response to the presence of unavoidable lateral pressure during the winding. The side plates were left on during heat treatment (HT). Stainless steel foil was placed between the $MgB_2$ wire leads and the Cu pads so that the wire would not diffusion bond to them during HT. The coil was HTed in a kiln using a SS retort with a flowing Ar. The peak temperature was 675°C applied for 30 minutes (see Figure 2), the ramp up time was 1.75 h and the ramp-down time was approximately 5-6 h. After HT the side plates were removed, and G-10 insulation was inserted between the Cu and plates and the SS flange, affixed with nylon screws. The strands occupied 49% of the coil pack. The $MgB_2$ area was 12.7% of the total winding area.

**Coil Epoxy Impregnation**

The coil was impregnated by placing it in a tray of vacuum-degassed mixed Stycast 1266 epoxy. During immersion of the coil a slight overpressure forced epoxy into the void spaces. After removal from the epoxy bath the coil curing was initiated by an insertion into a furnace at 65°C for 30 minutes, after which the coil was removed. Total curing time was estimated at 6 h.

**Fixture Mounting and Critical Current Measurement**

The coil was mounted on hangdown rods (a sister coil, readied for testing is shown in Figure 3) in preparation for insertion in the LHe cryostat. It was provided with voltage taps attached at layers 0, 4, 6, 8, 10, 12, and 16. The typical distance between successive taps was about 5 m (about twice the length per layer. Twisted wires (the potential leads) were attached to most of the taps and brought out to the cryostat head. Two heavy current leads were attached to the coil ends and also brought out to the cryostat head. After pre-cooling in $LN_2$ and final cooling to 4.2 K in LHe the system was connected up for a standard four-terminal I-V measurement. After setting the maximum take-off-voltage trigger, $V_{to}$, the current was slowly ramped up in a search for the normal-state transition at the critical current, $I_c$, corresponding to a voltage, $V_c < V_{to}$, based on the voltage criterion of 1 µV/cm. Values for $n$ were obtained by taking $E = \text{Const}(I/I_c)$ as the form for current and voltage above $I_c$, and using the slope of a log-log plot.

**Results**

The transport current testing results for various segments of the coil are displayed in Figure 4 (a) and (b). Critical current, $I_c$, defined using the 1 µV/cm criterion, is listed for all segments, as well as for the overall coil, in Table 1. Figure 4 (a) shows the raw data, in which a zero offset (but no baseline) was seen. These offsets seemed to be spurious in nature (perhaps inductive coupling with the voltage leads), and were removed in the close-up of Figure 4(b) to compare various coil segments. The overall coil $I_c$ is 127 A, with a variation over the segments of 123-147 A. The layers 4-6 seem to contain a small defect (indicated by the lower $I_c$ and the degraded $n$-value), otherwise the overall coil performance is fairly uniform. We note the curious fact that segment 0-4 and 4-6 seem to be developing a voltage in a small current interval preceding the point at which the whole coil does. The origin of this is unclear, the most likely reason is either some small leakage current, or some voltage gradient in the region of the whole-coil current and voltage taps (which are extended over about 4 cm). Room temperature measurements showed very high resistances between the conductor and the former (MΩs). Room temperature voltage measurements across the taps (with a current of 100 mA) gave voltages which were relatively consistent with the nominal voltage tap positions, with some variation probably due to Mg-Cu reaction induced variation in stabilizer resistivity.

Figure 4 data is replotted on a log-log scale in Figure 5. The *n*-values extracted are listed in Table 1; they range from 17-56 but are mostly (except for the 4-6 segment) greater than 30.

## Summary

A racetrack coil using $MgB_2$/Cu strand has been fabricated and tested for transport current density at 4.2 K and self field. The monofilamentary strand, 1.0 mm OD, was insulated with S-glass braid. Eighty turns of strand (42 m) were wound onto a stainless steel former with outer dimensions 25 cm x 10 cm x 0.86 cm. The coil was heat treated at 675°C for 30 minutes in flowing Ar. The strand, which had a superconducting fraction of 26%, occupied 49% of the total coil pack cross sectional area. The coil $I_c$ at 4.2 K and self field was 127 A (using a 1 µV/cm criterion), with an *n*-value of 41. Individual segments had $I_c$s ranging from 123 to 147 A, with *n*-values of from 17 to 56. The coil overall had a $J_{c,sc}$ of 6.12 x $10^4$ A/cm$^2$, a $J_e$ in the wire of 1.59 x $10^4$ A/cm$^2$, and a winding $J_e$ of 7.9 x $10^3$ A/cm$^2$ at 4.2 K in self field.

## Acknowledgements

This work was supported by the State of Ohio Technology Action Fund and NASA under contract No. NNC04CA41C.

**List of Tables**

Table 1. Coil Layers and Voltage tap positions.

| Layer | Layer Total, m | Running Total, m | Voltage Tap No. | $I_c$, A, for 1 μV/cm | $n$ |
|---|---|---|---|---|---|
| 0 | 0 | 0 | $V_0$ | | |
| 1 | 2.36 | 2.36 | | | |
| 2 | 1.50 | 3.87 | ($V_0$-$V_1$) | 124 | 38 |
| 3 | 3.43 | 7.29 | | | |
| 4 | 2.54 | 9.83 | $V_1$ | | |
| 5 | 2.54 | 12.37 | ($V_1$-$V_2$) | 125 | 17 |
| 6 | 2.64 | 15.01 | $V_2$ | | |
| 7 | 2.62 | 17.63 | ($V_2$-$V_3$) | 138 | 56 |
| 8 | 2.59 | 20.22 | $V_3$ | | |
| 9 | 2.74 | 22.96 | ($V_3$-$V_4$) | 147 | 53 |
| 10 | 2.82 | 25.78 | $V_4$ | | |
| 11 | 2.57 | 28.35 | ($V_4$-$V_5$) | 139 | 31 |
| 12 | 2.84 | 31.19 | $V_5$ | | |
| 13 | 3.25 | 34.44 | | | |
| 14 | 2.24 | 36.68 | ($V_5$-$V_6$) | 133 | 32 |
| 15 | 2.79 | 39.47 | | | |
| 16 | 2.87 | 42.34 | $V_{END}$ | | |
| whole | | 42.34 | ($V_0$-$V_{END}$) | 127 | 41 |

Table 2. Coil Parameters.

| Parameters | |
|---|---|
| Wire Designation | HTR 475 |
| Wire Type | $MgB_2$/Cu |
| Wire OD | 1.01 mm |
| Insulation | S-glass braid |
| Turns | 80 |
| Total length | 42 m |
| Coil pack cross section | 20.45 mm X 6.35 mm |
| Heat treat temperature | 675°C |
| Heat treat time | 30 min. |
| Superconductor fraction | 26% |
| Wire cross section as a % of the effective coil area | 49% |
| **Transport Results** | |
| $I_c$ of the coil, in self field, 4.2 K, 1 µV/cm | 120 A |
| $J_c$ of the wire in the coil, in self field, 4.2K | 61,200 A/cm$^2$ |
| $J_e$ of the wire in the coil, in self field 4.2 K | 15,900 A/cm$^2$ |
| $J_e$ for the effective coil area, in self field 4.2 K | 7,930 A/cm$^2$ |

# List of Figures

Figure 1. Schematic of coil former.

Figure 2. HT profile for coil.

Figure 3. Coil, mounting, and wiring.

Figure 4. Transport results for coil. (a) Raw data including offset. (b) offset corrected data. Voltage criterion at 1µV/cm are shown for 5 m segments (two layer segments) of the coil as well as the whole coil.

Figure 5. *n*-values for various coil segments.

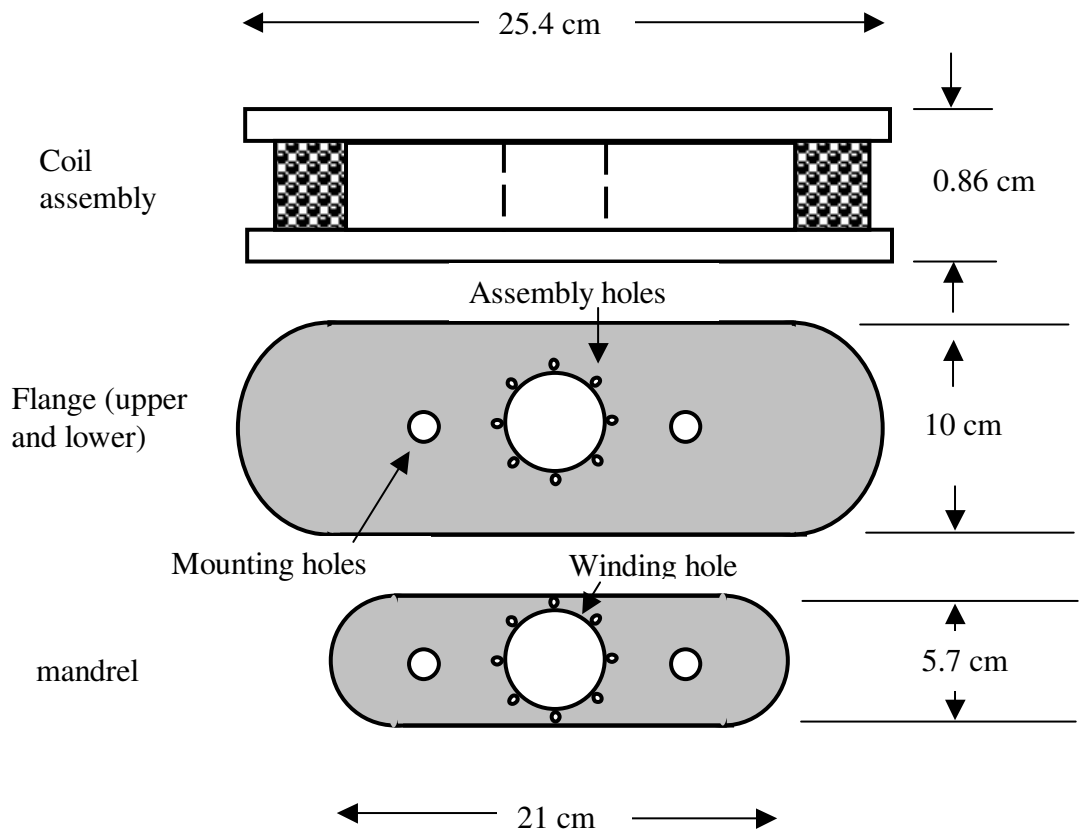

Figure 1.

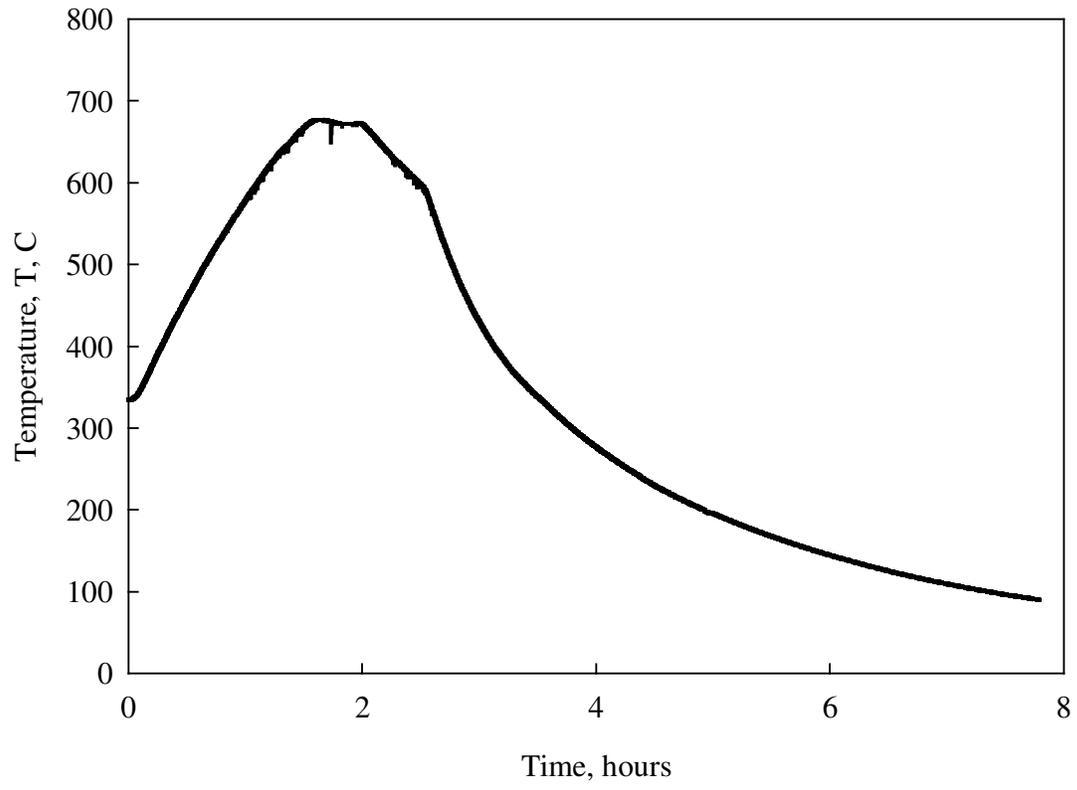

Figure 2.

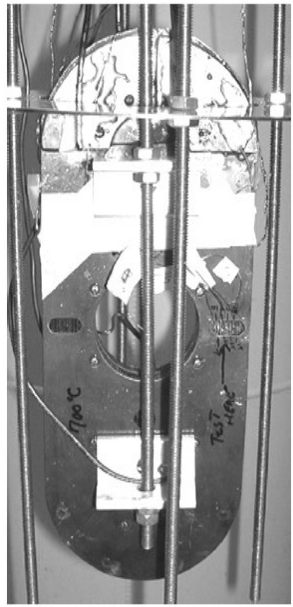

Figure 3.

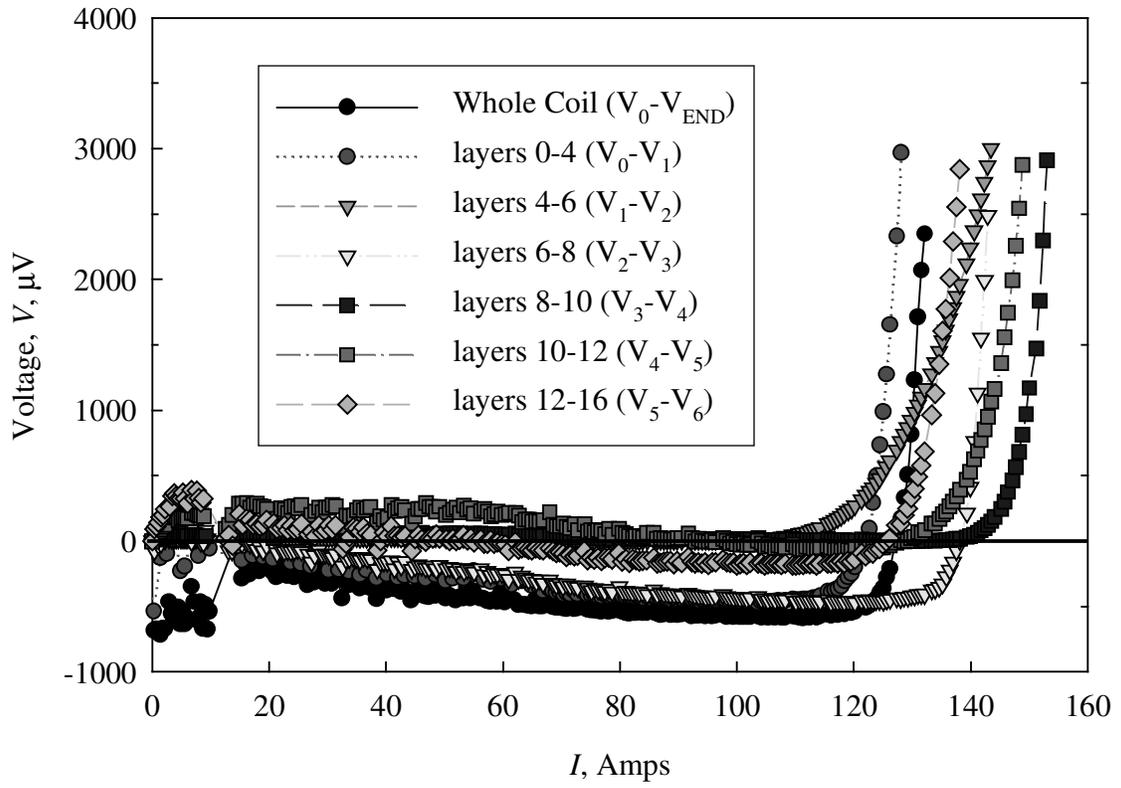

Figure 4(a)

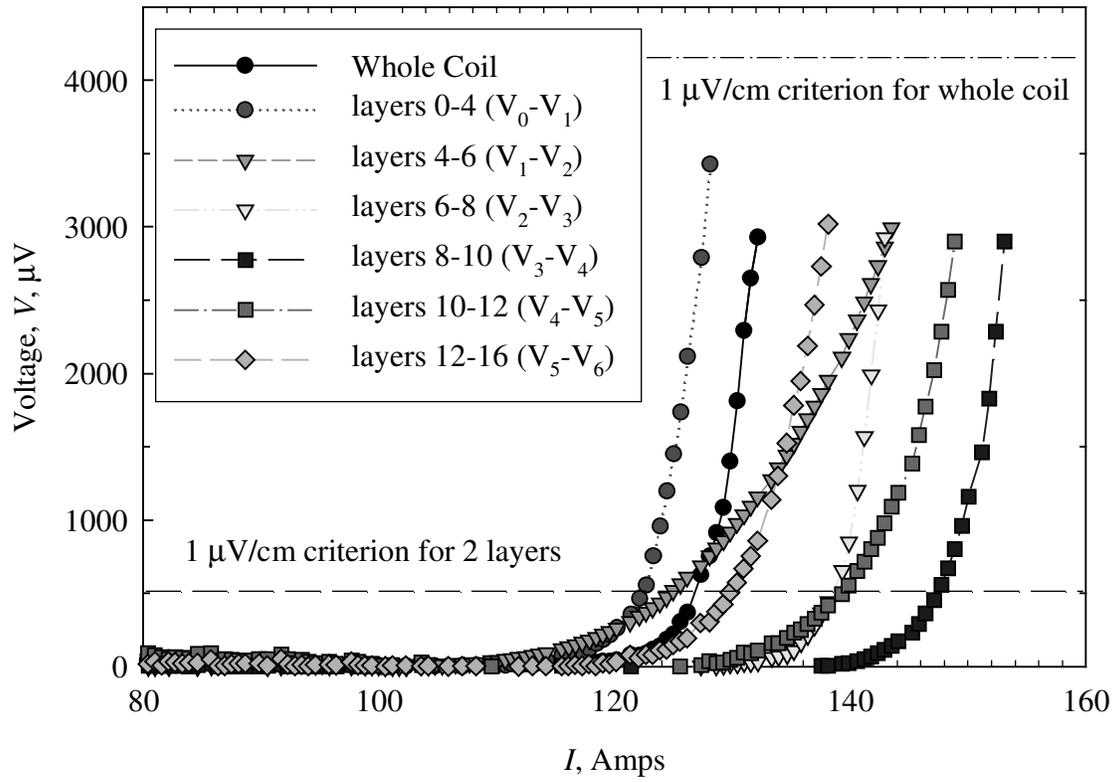

Figure 4 (b).

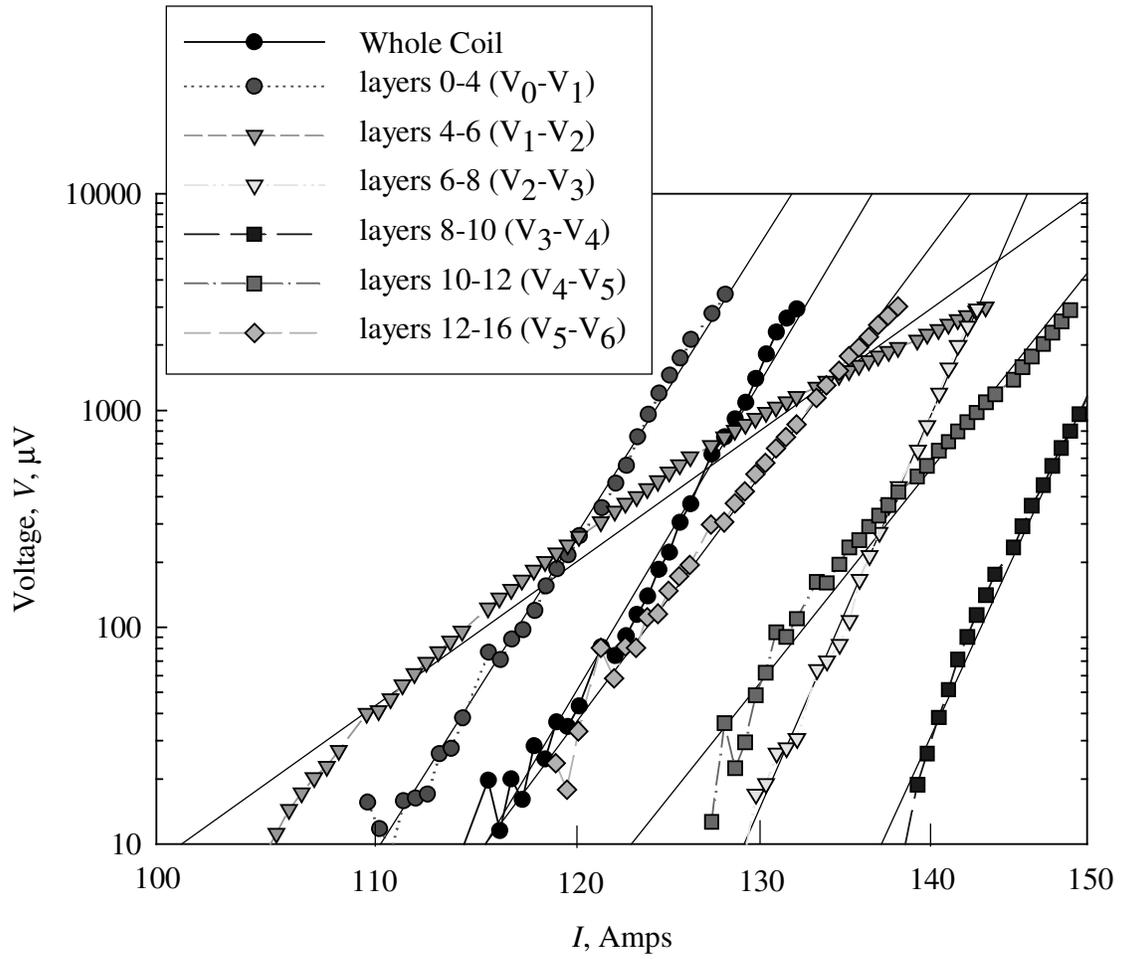

Figure 5.